\documentclass[prb,floatfix,twocolumn,showpacs,amsmath,amssymb]{revtex4}
\usepackage{graphicx}
\usepackage{dcolumn}
\usepackage{multirow}
\usepackage{bm}
\begin{document}

\title{Absence of static stripes in the two-dimensional $t{-}J$ model 
determined using an accurate and systematic quantum Monte Carlo approach}
\author{Wen-Jun Hu, Federico Becca, and Sandro Sorella}
\affiliation{Democritos Simulation Center CNR-IOM Istituto Officina dei 
Materiali and International School for Advanced Studies (SISSA), Via Bonomea 
265, 34136 Trieste, Italy}
\date{\today}

\begin{abstract}
We examine the two-dimensional $t{-}J$ model by using variational approach
combined with well established quantum Monte Carlo techniques [S. Sorella 
{\it et al.}, \prl {\bf 88}, 117002 (2002)] that are used to improve 
systematically the accuracy of the variational ansatz. Contrary to recent 
density-matrix renormalization group and projected entangled-pair state 
calculations [P. Corboz {\it et al.}, \prb {\bf 84}, 041108(R) (2011)],
a uniform phase is found for $J/t=0.4$, even when the calculation is biased 
with an ansatz that explicitly contains stripe order. Moreover, in the small 
hole doping regime, i.e., $\delta \lesssim 0.1$, our results support the 
coexistence of antiferromagnetism and superconductivity.
\end{abstract}


\maketitle

{\it Introduction.}
The comprehension of the low-energy properties of strongly-correlated systems
remains one of the biggest challenges in modern condensed matter physics.
Indeed, although a fair good understanding has been achieved in some limiting 
cases (especially for large spatial dimensions, thanks to dynamical mean-field
theory~\cite{georges,jarrell}), many important questions remain wide open in 
the two-dimensional case, where the competition between charge/spin ordering 
and superconductivity is very strong. Unfortunately, in this case, there are 
not unbiased techniques that may be used to obtain accurate results for low 
temperatures and large system sizes. Therefore, several approximate methods 
have been developed and applied in the last years, like for example variational 
(VMC)~\cite{gros} and fixed-node (FN) Monte Carlo,~\cite{ceperley} 
density-matrix renormalization group (DMRG)~\cite{dmrg} or its developments 
based upon the so-called tensor network states, including multi-scale 
entanglement renormalization ansatz (MERA)~\cite{MERA} and projected 
entangled-pair states (PEPS)~\cite{PEPS}, which has been recently generalized 
to fermionic systems~\cite{fermionicPEPS} and infinite lattices 
(iPEPS).~\cite{iPEPS}

Different calculations on the $t{-}J$ model have shown contradicting 
outcomes,~\cite{lin,putikka,manousakis,lee,calandra,white1,white2,lugas,spanu}
and the whole phase diagram of this model is still highly debated. One 
important issue, related to the mechanism of pairing in the cuprate 
materials, is whether some charge instability may take place (at $q=0$, leading
to phase separation, or at finite $q$, leading to the so-called stripes) or 
instead the homogeneous ground state is stable.~\cite{vojta}
In the latter case, the residual attraction among quasi-particles may lead to 
a superconducting state. Previous FN calculations emphasized the existence of 
a stable superconducting ground state,~\cite{dagotto} while DMRG and iPEPS 
results suggested a stripe order.~\cite{corboz}

The competition between superconductivity and stripes have been studied in 
several papers and different aspects have been addressed in the recent 
past.~\cite{ogata,raczkowski,capello} For example, two of us showed that a 
relatively small anisotropy in the super-exchange (and hopping) parameters may
lead to a striped order.~\cite{becca} In this regard, it is crucial to have a 
controlled method that may give variational results, in order to make a direct
comparison of energies (and other correlation functions) among different 
methods and reach a final consensus.

In this Rapid Communication, we adopt the same method used in 
Ref.~\onlinecite{dagotto}: by applying few Lanczos steps to the variational 
wave function and by filtering out its high-energy components (by means of the
Green's function Monte Carlo with the FN approximation), the accuracy of the 
calculations may be highly improved. This approach is particularly effective 
at low doping and is actually unbiased at half filling. Moreover, an estimation
of the exact energy may be given by the variance extrapolation: besides the 
energy, also the variance of the state can be calculated, and the energy with 
zero variance can be extracted. From our finding, the existence of a striped 
phase for $\delta \approx 1/8$ is rather unlikely: even the best approximation
to the ground state does not show any evidence towards charge inhomogeneity. 
Although the present calculations cannot rule out the possibility to have
small static stripes, our Monte Carlo approach is expected to reproduce 
qualitatively correct ground-state properties; in particular, it is reliable 
for determining the spatial dependent hole density: whenever an external 
modulated potential is added to the $t{-}J$ Hamiltonian, the FN approximation
gives rise to stripes, even when the initial state is chosen to be homogeneous.

{\it Model and methods.} 
The $t{-}J$ model on the two-dimensional square lattice is defined by:
\begin{equation}\label{eq:model}
{\cal H} = -t \sum_{\langle i,j \rangle \sigma} 
c_{i,\sigma}^\dag c_{j\sigma} + H.c. + 
J \sum_{\langle i,j \rangle}  \left ( {\bf S}_i \cdot {\bf S}_j - 
\frac{1}{4} n_i n_j \right ),
\end{equation}
where $\langle \cdots \rangle$ indicates nearest-neighbor sites, 
$c_{i,\sigma}^\dag$ ($c_{i,\sigma}$) creates (destroys) an electron with spin 
$\sigma$ on the site $i$; ${\bf S}_i$ and $n_i$ are the spin and density
operators on the site $i$, respectively. The $t{-}J$ Hamiltonian is defined in
the subspace without doubly occupied sites. In the following, we will take the 
amplitude for nearest-neighbor hopping $t=1$, and consider the super-exchange
$J/t=0.4$. The hole doping will be denoted by $\delta=1-N/L$, where $N$ and $L$
are the number of electrons and sites, respectively. Periodic boundary 
conditions are taken in both directions and $L \times L$ or 45-degree tilted 
lattices (with $L=2 l^2$, $l$ being an odd integer, so that the non-interacting
ground state is non-degenerate at half filling) are considered. 

Our starting variational wave function is defined as
\begin{equation}\label{eq:wavefunction}
|\Psi_v\rangle = {\cal P}_N {\cal P}_G {\cal J}_d {\cal J}_s |\Phi_{MF}\rangle,
\end{equation}
where ${\cal P}_N$ is the projector onto the subspace with $N$ electrons, 
${\cal P}_G$ is the Gutzwiller projector, which enforces no double occupation 
on each site; ${\cal J}_d=\exp(1/2 \sum_{i,j}u_{ij}n_i n_j)$ and 
${\cal J}_s=\exp(1/2 \sum_{i,j}v_{ij}S^z_i S^z_j)$ are density-density and 
spin-spin Jastrow factors, respectively. Finally, $|\Phi_{MF}\rangle$ is a 
mean-field state that may contain BCS pairing, antiferromagnetic order, 
or both. In our recent papers,~\cite{lugas,spanu} we have shown that very  
good variational energies can be obtained by orienting the magnetic order 
parameter in the $x{-}y$ plane, so that quantum fluctuations may be included 
thanks to the Jastrow term ${\cal J}_s$. In this case, however, the wave 
function takes the form of a Pfaffian.~\cite{lugas,spanu} Conversely, whenever
the antiferromagnetic order is taken along the $z$ direction, we deal with a 
determinant.~\cite{gros} The variational parameters are the $u_{ij}$'s and 
$v_{ij}$'s (for all independent distances in the lattice) and few parameters 
that describe the mean-field state $|\Phi_{MF}\rangle$ (i.e., the pairing 
amplitude $\Delta_{BCS}$, the antiferromagnetic parameter $\Delta_{AF}$, 
as well as the chemical potential and the next-nearest-neighbor hopping
describing  the variational electron dispersion). Due to the presence of strong
correlations (i.e., the Gutzwiller projector and the Jastrow factors), 
a variational Monte Carlo approach is required to compute the energy and all 
physical observables.

The accuracy of the wave function~(\ref{eq:wavefunction}) may be improved in 
different ways. The first one is by applying Lanczos steps:
\begin{equation}\label{eq:lanczos}
|\Psi_p\rangle = (1+\sum_{k=1}^{p}\alpha_k H^k)|\Psi_v\rangle,
\end{equation}
where the $\alpha_k$'s are additional variational parameters. Clearly,
whenever $|\Psi_v\rangle$ is not orthogonal to the exact ground state,
$|\Psi_p\rangle$ converges to it for large $p$. However, on large sizes,
only few steps can be efficiently performed: here, we consider the case with 
$p=1$ and $p=2$ ($p=0$ corresponds to the original variational wave function). 
Moreover, an estimation of the true ground-state energy may be achieved by the
variance extrapolation: for sufficiently accurate states, we have that
$E \approx E_{\rm ex} + {\rm const} \times \sigma^2$, where
$E=\langle{\cal H}\rangle/L$ and 
$\sigma^2=(\langle{\cal H}^2\rangle-\langle{\cal H}\rangle^2)/L$
are the energy and variance per site, respectively. Therefore, the exact
ground-state energy $E_{\rm ex}$ may be assessed by fitting $E$ vs $\sigma^2$
for $p=0,1$, and $2$.

\begin{figure}
\includegraphics[clip,width=9cm,height=7cm]{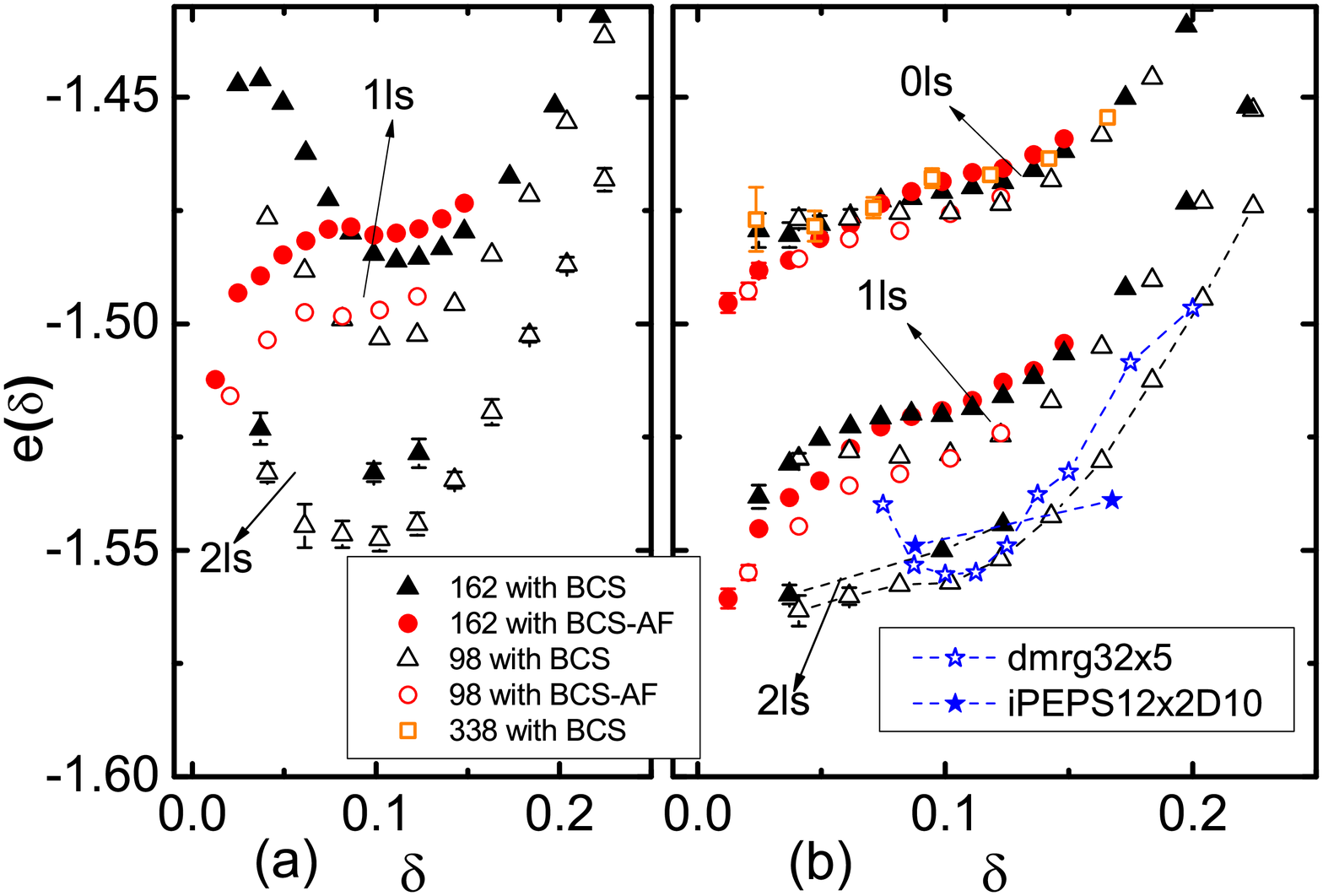}
\caption{\label{fig:energy}
(Color online). Energy per hole as a function of the doping for $J/t=0.4$.
Variational (left) and fixed-node (right) results are reported for $p=0$ and 
$1$ ($p=0,1$ and $2$) Lanczos steps for the wave function with (without)
antiferromagnetism. The best variational DMRG and iPEPS energies~\cite{corboz}
and the fixed-node with $p=2$ are connected by dashed lines for a better 
comparison.}
\end{figure}

Another way to improve the VMC calculations is through the FN 
approach,~\cite{ceperley} where the ground state of an auxiliary FN Hamiltonian
is obtained. In this case, the main approximation relies on the fact that the 
nodal surface is assigned {\it a priori}, by taking a given guiding function 
that is usually the best variational state. Most importantly, the resulting 
energies are still variational, so to have a totally controlled approximation 
of the original problem.~\cite{ceperley} 

In this paper, the guiding function is obtained by optimizing the Jastrow and 
the mean-field state, with the method described in 
Ref.~\onlinecite{sorella}. Then, we find the best Lanczos parameters $\alpha_p$
for $|\Psi_p\rangle$; finally, we perform the FN calculations with $p=0,1$, 
and $2$.

\begin{figure}
\includegraphics[clip,width=9cm,height=7cm]{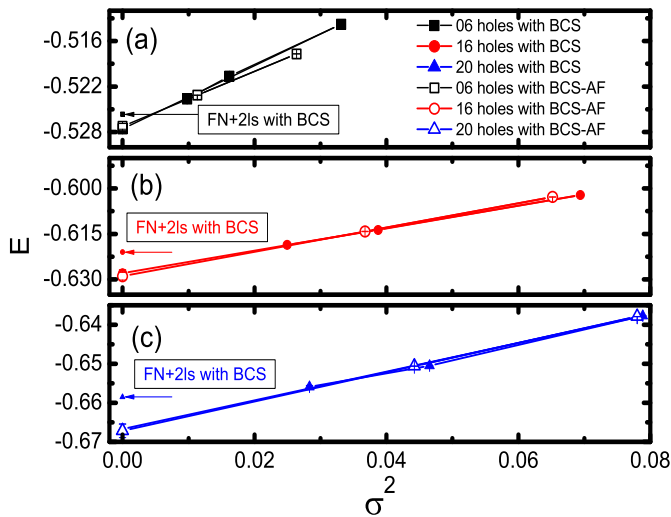}
\caption{\label{fig:extrapol}
(Color online) Variational results for the variance extrapolation on a 162-site
cluster, for different numbers of holes: $p=0$ and $1$ ($p=0,1$ and $2$) 
Lanczos steps have been performed on the wave function with (without) 
antiferromagnetism. The best fixed-node results are also marked by arrows.}
\end{figure}

{\it Results.}
Before showing the results on large systems, we would like to mention that
a very good accuracy on small lattices (where Lanczos diagonalizations can be 
performed) is obtained. We compared our results with the exact ones on the
26-site lattice for 2 and 4 holes, and different values of $J/t$ (see 
supplementary material~\cite{supplementary}). Both the Lanczos and the FN 
techniques largely improve the variational wave function and the best FN
calculations (with 2 Lanczos steps) reaches an accuracy of 
$(E_{\rm ex}-E)/E_{\rm ex} \approx 0.002$ and $\approx 0.003$ for 2 and 4 
holes, respectively (for $J/t=0.4$).

Let us now move to larger sizes and first analyze the tendency towards phase 
separation. In Fig.~\ref{fig:energy}, we show our results of the energy per 
hole $e(\delta)=[E(\delta)-E(0)]/\delta$ for various cluster sizes.~\cite{note}
$e(\delta)$ is a powerful detector for phase separation: a monotonic behavior
of $e(\delta)$ vs $\delta$ indicates a finite compressibility and a stable 
uniform phase, while a minimum, on finite systems, or a flat behavior in the
thermodynamic limit, indicate an instability.~\cite{lin} Close to half filling,
the Pfaffian wave function is considerably better than the simple 
superconducting state, clearly indicating a coexistence of pairing and 
antiferromagnetic order.~\cite{lugas,spanu} As the doping increases,
the antiferromagnetic parameter decreases and eventually vanishes for 
$\delta \approx 0.1$. The general trend is clear: the increased accuracy of 
the calculation favors the homogeneous state, marked by a monotonic behavior 
of the energy per hole vs the doping. In particular, one Lanczos step strongly 
improves the quality of the results, the gain in the FN energy being 
approximately $0.05t$, independently of $\delta$. Even the second Lanczos step 
is efficient for these large sizes, providing a further energy gain of about 
$0.02t$. We also mention that the results obtained with the variance 
extrapolation are consistent with the DMRG and iPEPS ones;~\cite{corboz} 
indeed, we have that $e(\delta)=-1.61(1)$ for 
$0.03 \lesssim \delta \lesssim 0.12$. Remarkably, we have obtained the same 
extrapolated values (within three error-bars) by using the two wave functions 
with or without antiferromagnetic order, see Fig.~\ref{fig:extrapol}. However,
the extrapolated values have too large error-bars and cannot be used to study
the issue of phase separation.

The application of few Lanczos steps on a given wave function is not size
consistent; nevertheless, an estimation of the thermodynamic limit can be 
attempted by considering the largest size, where the $p=0$ calculations do 
not show significant size effects. Therefore, we have considered $p=2$ FN
calculations for $L=162$ (or even $98$ for $\delta \gtrsim 0.17$), which 
compare well with the best energies obtained by DMRG and iPEPS. The latter 
ones provide  slightly more accurate energies for $\delta \simeq 0.1$. However,
considering that all these methods are significantly away from the estimated 
exact energy per hole obtained by DMRG and variance extrapolations 
(i.e., $e(\delta) \simeq -1.61$), this difference looks essentially irrelevant.
In contrast with DMRG and iPEPS that find a minimum in the energy per 
hole,~\cite{corboz} our best FN approximations do not show any tendency to 
phase separation for any doping, and, therefore, represent a thermodynamically
stable phase corresponding to a well defined variational state.

\begin{figure}
\includegraphics[clip,width=9cm,height=7cm]{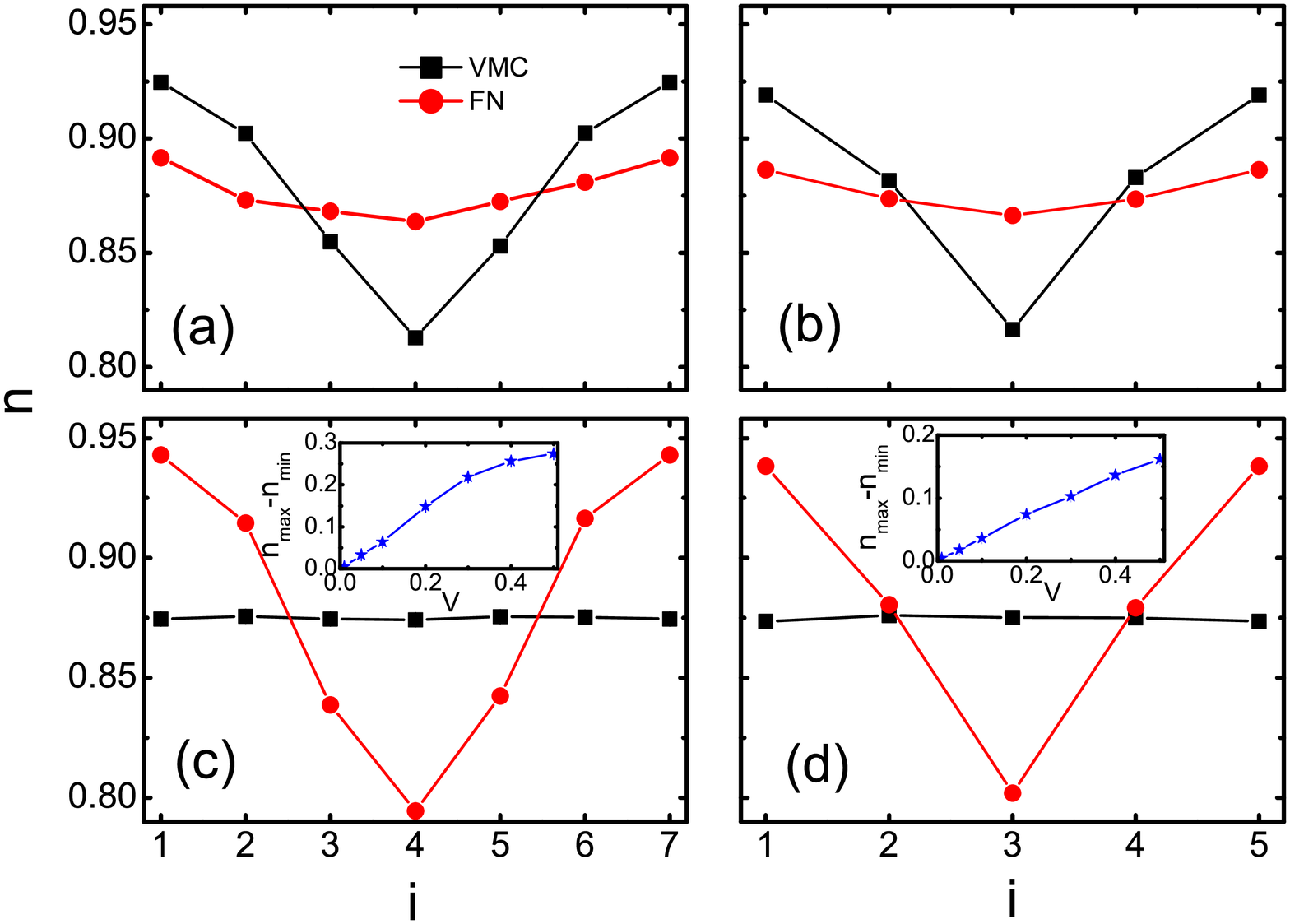}
\caption{\label{fig:density}
(Color online) Upper panels: local density $n_i$ when a site-dependent chemical
potential with $\delta \mu=1.6$ [see Eq.~(\ref{eq:chempot})] is added to the 
variational wave function; the cases with $l_s=12$ (a) and $8$ (b) are 
reported. Lower panels: local density $n_i$ when a site-dependent potential
[see Eq.~(\ref{eq:potential})] is added to the $t{-}J$ Hamiltonian, with 
$l_s=12$ and $V=0.2$ (c) and $l_s=8$ and $V=0.4$ (d). Variational and 
fixed-node results are reported for a $12 \times 12$ cluster and $\delta=1/8$.
Insets: the difference between the largest and the smallest local density 
(at the fixed-node level) as a function of $V$.}
\end{figure}

\begin{figure}
\includegraphics[clip,width=8cm,height=7cm]{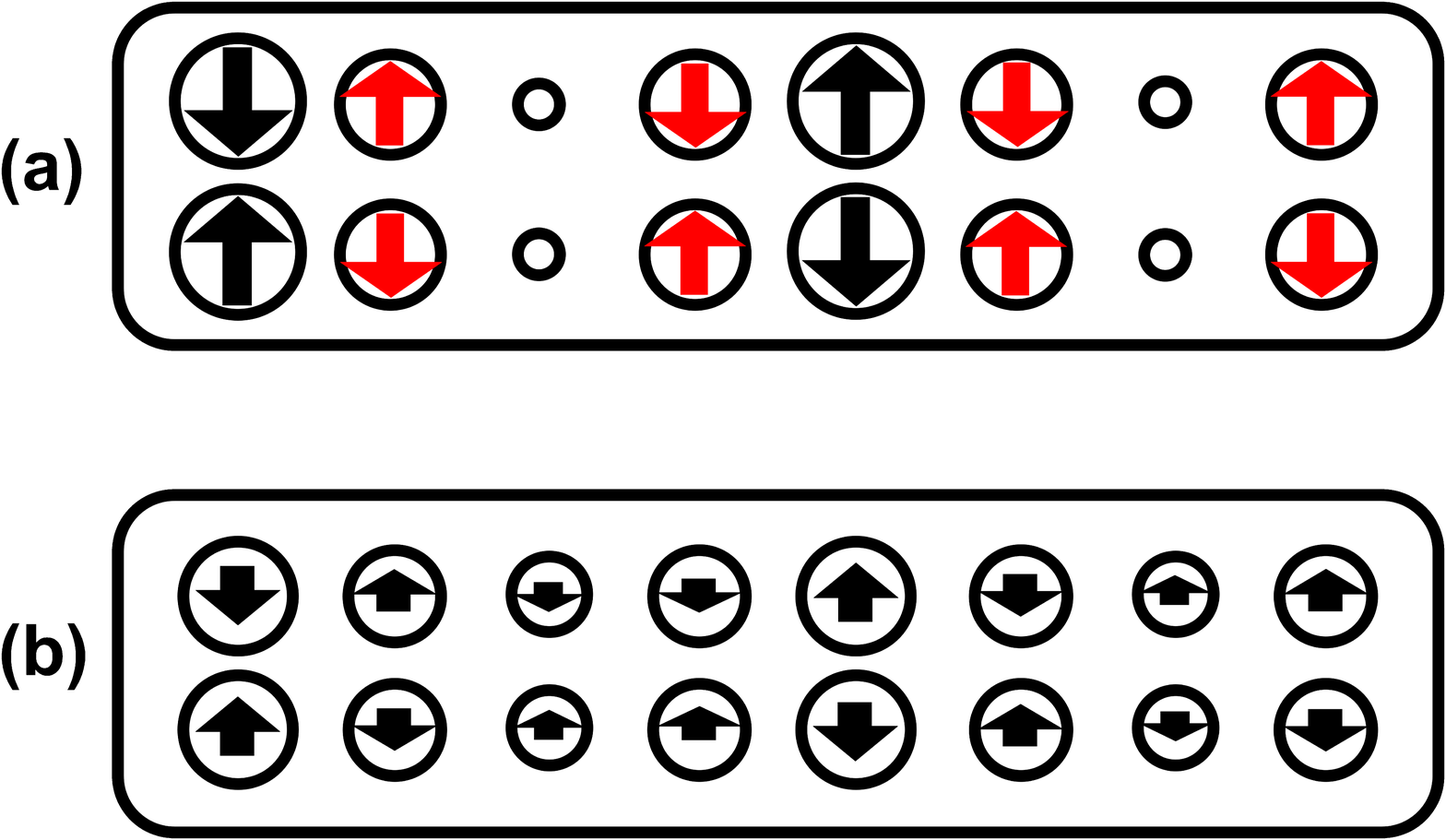}
\caption{\label{fig:stripe}
(Color online) Initial variational ansatz with stripe order (a) and fixed-node 
calculation (b) for charge and spin distributions in the $2 \times 8$ unit cell
of a $16 \times 16$ lattice. The size of the circles and arrows is proportional
to the electron density and spin along $z$, respectively. Largest symbols in 
the variational calculations: $\langle n_{R_i} \rangle=0.92$, 
$\langle S^z_{R_i}\rangle=\pm 0.09$.}
\end{figure}

Let us now consider the more subtle issue of stripes. Recently, DMRG and iPEPS
calculations suggested that the ground state has charge (and spin) modulations,
at least close to $\delta=1/8$.~\cite{corboz} Up to now, we have considered 
a uniform mean-field state $|\Phi_{MF}\rangle$, clearly biasing the VMC results
towards a homogeneous state. Despite the fact that the FN method can in 
principle remove this bias and give rise to non-uniform results, we have not 
found any evidence in favor of stripes with this variational ansatz. 

In order to gain some evidence that a charge inhomogeneity is not stabilized 
in the low-doping regime, we add a site-dependent chemical potential in the 
mean-field Hamiltonian
\begin{equation}\label{eq:chempot}
\mu_{R_i} = \mu_0 + \delta \mu \cos \left (\frac{4\pi}{l_s} x_i \right )
\end{equation}
where $R_i=(x_i,y_i)$ is the coordinate of the site $i$ and $l_s$ equal to 8 
or 12. By starting from a finite $\delta \mu$, the VMC optimization leads to
a perfectly uniform state with $\delta \mu=0$; moreover, FN calculation 
strongly reduces the density modulation present in the original variational 
wave function, see Fig.~\ref{fig:density}. Although a small inhomogeneity 
remains in the density profile, the FN energy is always higher than the one 
with $\delta \mu=0$. For these calculations, we considered $12 \times 12$, 
$16 \times 16$, and $24 \times 24$ lattices and $\delta=1/8$. Similar results 
have been obtained also for $\delta=1/12$ on a $12 \times 12$ lattice (not 
shown).

In order to show the effectiveness and the reliability of the FN method to
to detect charge inhomogeneities, we add a modulated potential directly in the
$t{-}J$ Hamiltonian:
\begin{equation}\label{eq:potential}
V_{R_i} = V \cos \left ( \frac{4\pi}{l_s} x_i \right ).
\end{equation}
Then, we consider a uniform mean-field wave function and compute the local
density for $12 \times 12$ and $24 \times 24$ lattices and $\delta=1/8$. 
The results are also reported in Figs.~\ref{fig:density}. Clearly, the VMC 
results show a completely flat behavior of the density in different sites; 
by contrast, the FN simulations are able to recover a strongly modulated 
density. This fact demonstrates that the presence of charge order could be 
detected by using this approach, even when a uniform guiding function is used
in the FN technique.

Finally, we can also add a spin structure to the charge modulation, so to have:
\begin{eqnarray}
\langle n_{R_i}\rangle &=& (1-\delta) - 
\delta n \cos \left (\frac{4\pi}{l_s} x_i \right ) \\
\langle S^z_{R_i}\rangle &=& 
\delta s (-1)^{R_i} \sin \left (\frac{2\pi}{l_s} x_i \right ). 
\end{eqnarray}
The above structure implies a $2 \times l_s$ unit cell and contains the 
so-called $\pi$-shift, namely anti-parallel spins across the hole-rich sites 
at $x_i=0$ and $l_s/2$. In the following, we consider suitable variational 
parameters inside the mean-field Hamiltonian that defines the uncorrelated 
state (i.e., local chemical potentials and local magnetic fields),
such to reproduce a stripe with $l_s=8$ and take $\delta=1/8$ on a 
$16 \times 16$ lattice. Then, we optimize {\it all} parameters (for each site
independently) and observe that the initial stripe melts and a perfect uniform
state is finally recovered. Moreover, by performing the FN approach starting
from a variational state with stripe order, we always obtain that the charge 
and spin modulations are reduced and a much more uniform state is found,
see Fig.~\ref{fig:stripe}; we also notice that the $\pi$-shift is replaced by 
a small defect in a weak antiferromagnetic background.

{\it Conclusions}.
In this work, we have shown that the FN approach is particularly reliable, not
only to improve the energy of a given variational ansatz, but also to determine
the density profile of the ground state, in a way that is rather independent 
of the original ansatz. Indeed, the approximate FN ground state 
$|\Psi_{FN}\rangle$ is not a ``brute force'' variational ansatz, but it 
represents the ground state of a physical Hamiltonian that is different from 
the exact one only in the region where the variational wave function is close
to zero (namely within the so-called {\it nodal region}). Operators ${\cal O}$ 
that are diagonal in configuration space $|x\rangle$ (e.g., related to stripes
or antiferromagnetic order) are weakly affected by this nodal error. Indeed, 
in the expectation value of ${\cal O}$, which takes the form of 
$\sum_x \Psi_{FN}^2(x) {\cal O}_x$, the nodal region, where 
$\Psi_{FN}(x) \simeq 0$, provides a very little contribution, thus explaining 
the reliability of the FN approach.   

We have shown that the FN Monte Carlo, when combined with few Lanczos steps, 
is competitive with recent DMRG and iPEPS calculations, as far as the 
variational energy is concerned. The main outcome is that the ground state is 
homogeneous. No evidence of stripes are detected around $\delta=1/8$: at low 
doping, a uniform state is stabilized, containing both superconductivity and 
antiferromagnetism. Despite our findings, we have to conclude honestly that 
the low-doping phase diagram of the $t{-}J$ model is not settled yet, since
very accurate methods provide very different phases with almost comparable 
energies. We believe that future calculations that employ the FN approach
on top of iPEPS or DMRG may be helpful for the final 
understanding.~\cite{ducroo}

We thank P. Corboz and S.R. White for useful discussions and for providing us
with the DMRG and iPEPS results.


\begin{thebibliography}{99}
\bibitem{note} The energy at half filling has been computed for each cluster 
   independently. We have seen that this choice minimizes finite size effects,
   that become neglegible for $L \simeq 162$ sites.
\end{thebibliography}
\end{document}